\documentclass[12pt,preprint]{aastex}

\slugcomment{Submitted to Astrophysical Journal}
\shorttitle{Fermionic dark matter}
\shortauthors{Nakajima}

\begin{document}

\title{Fully Degenerate Self-Gravitating Fermionic Dark Matter: \\
Implications to the Density Profile of the Cluster of Galaxies A1689, and the Mass Hierarchy of Black Holes}

\author{Tadashi Nakajima}
\affil{National Astronomical Observatory of Japan \\ 
Osawa 2-21-1, Mitaka, 181-8588, 
Japan}

\and 

\author{Masahiro Morikawa}
\affil{Department of Physics, Ochanomizu University \\
2-1-1 Otsuka, Bunkyo, Tokyo,112-8610, Japan
}

\begin{abstract}
Equilibrium configurations of weakly interacting fully degenerate fermionic
dark matter are considered at various scales in the Universe. We treat the
general situations for the gravity from Newtonian to general relativity and
the degeneracy from nonrelativistic to relativistic. The formulation of the
problem is exactly the same as the case treated by Oppenheimer and Volkoff
in their paper on neutron stellar cores.
A dimensionless equilibrium configuration is specified by
a single parameter regardless of particle properties, the
Fermi velocity at the center, and the scalings of mass and
length are specified by
the rest mass and statistical weight
of the dark matter particle. 
We focus our 
attention to the flat-top nature of the mass column density profile
of the cluster of galaxies, A1689,
recently reported by Broadhurst et al. using gravitational lensing.
We convert the column density profile to a volume density profile
assuming spherical symmetry and derive a 3D encircled mass profile
of A1689, which is compared with the model profiles of degenerate fermion
structures. The flat-top profile is reproduced. The corresponding
fermion mass ranges from 2 eV to 30 eV depending on the actual
scale of the degenerate structure. If massive neutrinos are
the dominant dark matter, the rest mass will be about
4.7 or 2.3 eV respectively for Majorana or Dirac neutrinos.
The mass and size of the degenerate structure are
$10^{14}M_\odot$ and 100 kpc for Majorana neutrinos, and
5$\times10^{14}M_\odot$ and 300 kpc for Dirac neutrinos.
If we identify the fermions as heavier
sterile neutrinos, they yield the characteristic mass hierarchy of black
holes; giant black hole at the center of  a galaxy and the intermediate mass
black holes. Thus we propose the possibility that the mass hierarchy of
fermions determines that of black holes in the Universe.
\end{abstract}


\section{Introduction}

\bigskip

Recent precision observations have revealed that the unknown dark matter
dominates the matter contents of the Universe. We wish to study the possible
dynamical
structures of their existence, especially in a universal form. If the dark
matter is in the form of ordinary thermal
gas, the structure and the dimension would
be strongly dependent on the initial conditions and the environment of the
expanding universe. On the other hand if the dark matter is almost
degenerate, we will naturally find a universal structure. Moreover, we would
like to know the possibility that the dark matter forms black holes.

To make the problem setting better defined, we start our consideration from
the typical structure made from ordinary baryonic matter, in which an
electron and a nucleon form the basic ingredient through the electromagnetic
force. In the total energy $E=\frac{1}{2}m_{e}v^{2}-(e^{2}/r)$, the second
term represents the attractive force making the system collapse and the
first term represents the pressure against it through the Heisenberg's
uncertainty principle. Actually putting the expression of the de Broglie wave
length $r=\hbar /(m_{e}v)$ in $E$, and extremizing it with respect to $r$,
we obtain the ground state bounding energy of hydrogen $E=-m_{e}e^{4}/(2\hbar
^{2})=-2.19\times 10^{-18}$ J and the Bohr radius $r=\hbar
^{2}/(e^{2}m_{e})=5.28\times 10^{-9}$ cm. The corresponding energy
density is $\rho =m_{p}/({\frac{4\pi }{3}}r^{3})=2.71$ g$\cdot$cm$^{-3}$. 
This is the
basic ingredient of the structure formed by electromagnetism. The
electron can collapse more. Setting $v\rightarrow c$ in de Broglie wave
length \noindent yields the Compton wave length. The corresponding energy
density becomes $\rho =m_{p}/({\frac{4\pi }{3}}r_{c}^{3})=6.93\times 10^{6}$
g$\cdot$cm$^{-3}$. This is the basic ingredient of white dwarfs. Further
collapse makes the electron and the proton into a neutron, through the inverse
beta decay process. Putting $m_{e}\rightarrow m_{p}$, we have $r=\hbar
/(m_{p}c)$, and therefore the corresponding energy density is $\rho
=m_{p}^{4}c^{3}/\hbar ^{3}=4.29\times 10^{16}$ g$\cdot$cm$^{-3}$. This is the
basic ingredient of neutron stars.

In general, the material formed by fully degenerate fermion of mass $m$
would have the energy density $\rho =m^{4}c^{3}/\hbar ^{3}$. A structure of
this density with radius $R$ has the mass $M=R^{3}\rho $. On the other hand
the size $R$ should be smaller than the limiting scale, Schwarzschild radius, 
$2GM\leq R$. This condition yields the maximum mass as

\begin{equation}
M_{fermi}=G^{-3/4}m^{-2}=m_{pl}^{3}/m^{2}  \label{Mfermi}
\end{equation}
where the Planck mass is defined as $m_{pl}=\left( \hbar c/G\right)
^{1/2}=2.18\times 10^{-5}$ g. Thus the quantum mechanics ($\hbar $),
gravity ($G$), and the particle physics ($m$) as well as relativity ($c$)
characterize the universal structure of a fully degenerate
fermion star (FDFS).

Incidentally, degenerate bosons form much smaller structures since
bosons have no exclusion principle like fermions. Therefore only the
Heisenberg uncertainty principle (quantum pressure) can support the structure
against the collapse due to gravity. The sole characteristic length scale is
the Compton wave length $l_{Compton}=\hbar /(mc)\approx R$, which must
be larger
than the limiting scale, Schwarzschild radius, $2GM\leq R$. This condition
yields the density $\rho =m^{2}m_{pl}^{2}c^{3}/\hbar ^{3}$ and 
\begin{equation}
M_{bose}=m_{pl}^{2}/m,
\end{equation}
which is known as the Kaup mass \citep{kaup}. 
This structure is a boson star. This
structure is smaller than that formed by fermions by a factor $m_{pl}/m$
and no further discussion will be given in this paper, apart from
a brief comment regarding the equation of state in \S2.

Now we proceed to consider the equilibrium
structures made of fermionic dark matter.
One may think that this problem is analogous to
the white dwarf case and the Chandrasekhar mass is 
the limiting mass. However this is not the right answer. In the case of a white
dwarf, the pressure is due to the degeneracy pressure of relativistic
electrons, but the gravity is Newtonian because it is due to the rest mass of
hadrons which are nonrelativistic. When we treat a degenerate star made
purely of dark matter itself, the total rest mass is also due to dark matter
particles themselves which can be relativistic. Therefore we need to
consider the gravitational mass of the star in a general relativistic
manner. This problem is therefore analogous to the case of a neutron star.
In the case of the neutron star, there is a complication due to the fact that
neutrons interact by nuclear forces. However in the case of weakly
interacting fermionic dark matter, it can be considered as ideal gas and the
equation of state is well defined. It turned out that this situation was
treated by the classical paper by \citet{OV39}, since they assumed free
neutrons and used the equation of state of ideal gas in their calculation.
Basically fully degenerate fermions can form very compact dense structures
including black holes \citep{bilic99,bilic03}.

On the other hand, from the observational side, we now have many
candidates of dense structures at various scales. The most massive
example is a 
huge dark matter distribution of the
cluster of galaxies, A1689, reported recently by
\citet{TB05a,TB05b}. They obtained the mass column density distribution of the
cluster of galaxies, A1689 by gravitational lensing.
The column density profile has a flat-top and we suspected that
this flat-top nature might be due to the degeneracy pressure of fermions.
We derive a volume density profile from the observed column density profile
assuming spherical symmetry and compare the observed 3D encircled mass
profile with our model profile of an FDFS.

The condense structure made from the fully degenerate fermion is not
restricted to a center of a cluster of galaxies. If we consider more massive
neutrinos, such as sterile neutrinos, the similar structures are realized in
scaled-down form as in Eq.(\ref{Mfermi}). If this structure is universal, 
we will find 
groups of black holes which have typical masses 
directly characterized by fermion masses.

The paper is organized as follows. 
The formalism by \citet{OV39} is introduced and
equilibrium solutions are discussed in \S2.
Readers who are not interested in the derivation of the solutions,
are advised to skip to \S2.3 where the properties of the solutions
are discussed.
The application of a nonrelativistic FDFS to
the cluster of galaxies A1689, is discussed in \S3.
The possible relation between the mass hierarchies of
black holes and sterile neutrinos are considered in \S4.

\section{Formalism}

Here we review the derivation of the general relativistic equilibrium
equations following \citet{RT34} and \citet{OV39}.

\subsection{Relativistic treatment of equilibrium}

The most general static
line element exhibiting spherical symmetry may be expressed in the form

\begin{equation}
ds^{2}=-e^{\lambda }dr^{2}-r^{2}d\theta ^{2}-r^{2}\sin ^{2}\theta d\phi
^{2}+e^{\nu }dt^{2},\quad \lambda =\lambda (r),\quad \nu =\nu (r).
\label{line_element}
\end{equation}

If the matter supports no traverse stresses and has no mass motion, then its
energy momentum tenor is given by

\begin{equation}
T_{1}^{1}=T_{2}^{2}=T_{3}^{3}=-p,\quad T_{4}^{4}=\epsilon,  \label{Ttensor}
\end{equation}
where $p$ and $\epsilon $ are respectively the pressure and the macroscopic
energy density measured in proper coordinates. Einstein's field equations
without the cosmological constant reduce to

\begin{equation}
8 \pi p = e^{-\lambda}\left(\frac{\nu^\prime}{r}+\frac{1}{r^2}\right)
 - \frac{1}{r^2},
\label{fieldeq1}
\end{equation}

\begin{equation}
8 \pi \epsilon = e^{-\lambda}\left(\frac{\lambda^\prime}{r}-\frac{1}{r^2}
\right) + \frac{1}{r^2},  \label{fieldeq2}
\end{equation}

\begin{equation}
\frac{dp}{dr}=-\frac{p+\epsilon }{2}\nu ^{\prime },  \label{fieldeq3}
\end{equation}
where primes denote differentiation with respect to $r$. These three
equations together with the equation of state of the material $\epsilon $ = $
p(\epsilon )$ determine the mechanical equilibrium of the matter distribution
as well as the dependence of the metric $g_{\mu \nu }$'s on $r$.

The boundary of the matter distribution is the value of $r=r_{b}$ for which $%
p=0$, and such that for $r<r_{b},p>0$. For $r<r_{b}$ the solution depends on
the equation of state of the material connecting $p$ and $\epsilon $. For many
equations of state a sharp boundary exists with a finite value of $r_{b}$.

In the empty space, $p = \epsilon = 0$,
surrounding the spherically symmetric distribution of matter,
the Schwarzschild's exterior solution is obtained:

\begin{equation}
e^{-\lambda (r)}=1-\frac{2m}{r},\quad e^{\nu (r)}=1-\frac{2m}{r},
\end{equation}
where $m$ is the Newtonian mass of the matter as calculated by a distant
observer.

Inside the boundary, Eqs. (\ref{fieldeq1}), (\ref{fieldeq2}), 
and (\ref{fieldeq3})
may be rewritten as follows. 
Using the equation of state $\epsilon = \epsilon(p)$,
Eq. (\ref{fieldeq3}) may be immediately integrated

\begin{equation}
\nu(r) = \nu(r_b) - \int^{p(r)}_0 \frac{2dp}{p + \epsilon(p)}, \quad
e^{\nu(r)} =
e^{\nu(r_b)} \exp\left[-\int^{p(r)}_0 \frac{2dp}{p+\epsilon(p)}\right].
\end{equation}

The constant $e^{\nu(r_b)}$ is determined by making $e^\nu$ continuous across
the boundary.

\begin{equation}
e^{\nu(r)} = (1-\frac{2m}{r})\exp\left[-\int^{p(r)}_0 \frac{2dp}{%
p+\epsilon(p)}\right].  \label{enu}
\end{equation}

Thus $e^\nu$ is known as a function of $r$ if $p$ is known as a function of $%
r$. Further in Eq.(\ref{fieldeq2}). introduce a new variable

\begin{equation}
u(r)=\frac{1}{2}r(1-e^{\lambda })\quad 
{\hskip 5mm}{\rm or}\quad e^{\lambda }=1-\frac{2u}{r}.  \label{equ}
\end{equation}

Then Eq. (\ref{fieldeq1}) becomes:

\begin{equation}
\frac{du}{dr} = 4\pi\epsilon(p)r^2.  \label{dudr}
\end{equation}

In Eq. (\ref{fieldeq1})
we replace $e^{-\lambda}$ by its expression (\ref{equ}) and 
$\nu^\prime$ by its expression (\ref{fieldeq3}). Then it becomes

\begin{equation}
\frac{dp}{dr} = - \frac{p+\epsilon(p)}{r(r-2u)}[4\pi p r^3 + u].  \label{dpdr}
\end{equation}

Eqs. (\ref{dudr}) and (\ref{dpdr}) form a system of two first-order
equations in $u$ and $p$. Starting with some initial values 
$u = u_0$ and $p=p_0$ at $r=0$, the 
two equations are integrated simultaneously to
the value $r=r_b$ where $p=0$, i.e., until the boundary of the matter
distribution is reached. The value of $u=u_b$ at $r=r_b$ determines the
value of $e^{\lambda(r_b)}$ at the boundary to the exterior solution, making

\begin{equation}
u_b = \frac{r_b}{2}[1-e^{-\lambda(r_b)}]=\frac{r_b}{2}
\left[1-(1-\frac{2m}{r_b})\right] = m.
\end{equation}

Thus the mass of this spherical distribution of matter as measured by a
distant observer is given by the value $u_b$ of $u$ at $r=r_b$.

The following restrictions must be made on the choice of $p_0$ and $u_0$,
the initial values of $p$ and $u$ at $r=0$:

(a) In accordance with its physical meaning as pressure, $p_0 \geq 0$.

(b) From Eq.(\ref{equ}) it is seen that for all finite values of $%
e^{-\lambda} $, $u_0 = 0$. Since $g_{11} = -e^\lambda$ must never be
positive, $u_0 \leq 0 $ for infinite values of $e^\lambda$ at the origin.
However, it may be shown that of all the finite values of $p_0$ at the
origin $p_0=0$ is the only one compatible with a negative value of $u_0$,
and that for equations of state of the type occurring in this problem even
this possibility is excluded, so that $u_0$ must vanish.

This can be seen from the
following argument. Having chosen some particular value of $p_0$ one may
usually represent the equation of state in that pressure range by $\epsilon = C
p^s$ with some appropriate value of $s$. Using this equation of state and
taking the approximate form of Eq (\ref{dpdr}) near the origin for the case $%
u_0 < 0$, and finite $p_0$, one obtains:

\begin{equation}
\frac{dp}{dr} = \frac{p + \epsilon(p)}{2r} = \frac{p + C p^s}{2r}.
\end{equation}

Integration of this equation shows that for $s<1$, $p_0 \geq 0$ can not be
satisfied, and for $s \geq 1$ only the value $p_0 = 0$ is possible. This
immediately excludes the possibility that degenerate bosons form an
equilibrium structure, because $p$ is independent of $\epsilon$,
if $p$ is solely due to thermal bosons \citep{landau}.
As we mentioned in \S1, a boson star is supported by
the quantum pressure of  ground-state bosons,
which is outside of the scope of the consideration of the equation
of state \citep{kaup}.
For the
equations of state used for degenerate fermions, always $s<1$ holds. It is
also be noted that the above equation together with Eq. (\ref{enu}) 
show that $%
e^{\nu(r)} \rightarrow \infty$ as $r \rightarrow 0$.

(c) A special investigation for any particular equation of state must be
made to see whether solutions exist in which $0 \leq u_0 \leq -\infty$ and $%
p \rightarrow \infty$ as $r \rightarrow 0$.

\subsection{Equation of state for degenerate Fermi gas}

If the matter consists of fermions of rest mass $\mu_0$ and 
statistical weight
$g$, and their thermal energy and all forces between them are
neglected, then a parametric form for the equation of state  \citep{landau}
is,

\begin{equation}
\epsilon = K (\sinh t - t),
\end{equation}

\begin{equation}
p=\frac{1}{3}K\left(\sinh t-8\sinh \frac{t}{2}+3t\right),
\end{equation}
where

\begin{equation}
K=\frac{\pi g\mu _{0}^{4}c^{5}}{8h^{3}},
\end{equation}
and

\begin{equation}
t=4\mathrm{arcsinh}\frac{p_{F}}{\mu _{0}c},
\end{equation}
where $p_{F}$ is the maximum momentum in the Fermi distribution and is
related to the proper particle density $N/V$ by

\begin{equation}
\frac{N}{V} = \frac{4\pi g}{3h^3} p_F^3 =\frac{4\pi g}{3}(\frac{\mu_0 c}{h})^3
\sinh^3\frac{t}{4}.
\label{ndensity}
\end{equation}

If we define the Fermi velocity $v_F$ by

\begin{equation}
\frac{p_F}{\mu_{0}c} = \frac{v_F}{\sqrt{1-(v_F/c)^2}},
\label{vf}
\end{equation}

\begin{equation}
t = 2 \log \left(\frac{1+v_F/c}{1-v_F/c}\right) \hskip 5mm {\rm or}
\hskip 5mm v_F/c = \tanh\frac{t}{4}.
\label{t-vf}
\end{equation}

$t$ and $v_F$ are independent of particle properties.

Substituting the above expressions for $p$ and $\epsilon$ into Eqs. 
(\ref{dudr})
and (\ref{dpdr}) one obtains:

\begin{equation}
\frac{du}{dr} = 4\pi r^2 K (\sinh t - t),  \label{dudr2}
\end{equation}

and

\begin{eqnarray}
\frac{dt}{dr} & = & -4/[r(r-2u)] \nonumber \\
& & \times (\sinh t - 2 \sinh 1/2t)/(\cosh t - 4 \cosh 1/2t +3) \nonumber \\
& & \times \left[\frac{4\pi}{3}K r^3 (\sinh t - 8 \sinh 1/2t +3t) + u\right].
\label{dtdr}
\end{eqnarray}

These equations are to be integrated from the values $u=0, t=t_0$ at $r=0$
to $r=r_b$ where $t_b = 0$ (which makes $p=0$), and $u = u_b$.

So far, the equations are written in relativistic units, i.e., such that $c
= 1, G=1$. This determines the unit of time and the unit of mass in terms of
still arbitrary unit of length. The unit of length is now fixed by the
requirement that $K = 1/(4\pi)$. From the dimensional analysis of
Einstein's field equations, this requirement fixes the unit of length to be

\begin{equation}
a=\frac{1}{\pi }\left(\frac{2}{g}\right)^{1/2}
\left(\frac{h}{\mu _{0}c}\right)^{3/2}
\frac{c}{(\mu_{0}G)^{1/2}},  \label{length}
\end{equation}
and the unit of mass to be

\begin{equation}
b = \frac{c^2}{G} a = \frac{1}{\pi} \left(\frac{2}{g}\right)^{1/2} 
\left(\frac{h}{\mu_0 c}\right)^{3/2} 
\frac{c^3}{(\mu_0 G^3)^{1/2}}.  \label{mass}
\end{equation}

Eqs. (\ref{dudr}) and (\ref{dtdr}) written in a dimensionless form become:

\begin{equation}
\frac{du}{dr} = r^2 (\sinh t - t),  \label{dudr3}
\end{equation}

\begin{eqnarray}
\frac{dt}{dr} & = & -4/[r(r-2u)] \nonumber \\
& & \times (\sinh t - 2 \sinh 1/2t)(\cosh t - 4 \cosh 1/2t +3) \nonumber \\
& & \times \left[\frac{1}{3} r^3 (\sinh t - 8 \sinh 1/2t +3t) + u\right]. 
\label{dtdr3}
\end{eqnarray}

For a given $t_0$, Eqs. (\ref{dudr3}) and (\ref{dtdr3}) can be integrated.
Fixing $t_0$ is equivalent to fixing $v_{F0}$.
As long as dark matter particles are fully
degenerate, the solution describes the equilibrium between the gravity and
degeneracy pressure from Newtonian to general relativistic gravity and
from nonrelativistic $v_F$ to relativistic $v_F$. Therefore the solution for
a given $t_0$ (or $v_{F0}$) 
is independent of particle properties, while the units of
length and mass ($a$ and $b$) are fixed by the particle properties $\mu_0$
and $g$.

\subsection{Discussion on solutions}

The Eqs.(\ref{dudr3}) and (\ref{dtdr3}) are numerically integrated using the
fourth-order Runge-Kutta method. 
Apart from the gravitational mass $u$, there is another mass indicator,
the dimensionless total rest mass $y_b$, defined as

\begin{equation}
y_b = \int_0^{r_b} \frac{32}{3} \sinh^3\frac{t}{4}/\sqrt{(1-2u/r)} r^2 dr,
\end{equation}

which is the integral of the number density with the proper volume inside
the 
radius $r_b$.
$t_0$, $v_{F0}/c$, $u_b$, $y_b$, and $r_b$ are given
for $t_0 = 0.1 \sim 14.0$ in Table 1. At $t_0 = 0.1$,
degenerate particles are nonrelativistic ($v_{F0}/c = 0.025$),
while at $t_0 = 14$,
particles are extremely relativistic ($v_{F0} = 0.998$). 
In Fig. 1, the relation between 
the gravitational mass $u_b$ and 
outer radius $r_b$ is plotted. 
For $t_0\leq3$, $u_b$ is an increasing function of $t_0$,
while $r_b$ is a decreasing function of $t_0$.
The maximum of $u_b=0.0766$ is reached for $t_0 = 3$ and
$r=0.663$. This is the maximum stable solution for which
$v_{F0}/c=0.635$, which is modestly relativistic.
In Fig.2, the radial profiles of the energy density, $\epsilon$,
and pressure, $p$, are plotted. The contribution of $p$
is relatively small compared to $\epsilon$. This solution
is termed as quasi-Newtonian by \citet{OV39}. In the case of
self-gravity of degenerate fermions themselves, a stable
configuration never reaches an extremely relativistic situation
unlike the case of white dwarfs. Therefore 
the formula of the Chandrasekhar mass applied
to this case over-estimates the maximum stable mass
by one order of magnitude.
For $t_0>3.0$, there is no stable solution \citep{OV39}. 
$u_b$ decreases takes a minimum value 
at ($u_b$,$r_b$)=(0.0395,0.364) for $t_0=8.0$.
For a large $t_0$, ($u_b$,$r_b$) spirals to $(\sim0.041,\sim0.29)$.

The gravitational mass defect, $\Delta = u_b - y_b$ is
one measure of stability of a general relativistic equilibrium configuration.
In Fig.3, both $y_b$ and $u_b$ are plotted against $t_0$ for
$0.1 < t_0 < 14.0$. 
For a small $t_0$, $\Delta$ is negative and
gradually decreases and 
takes a minimum value, $\Delta=-0.0029$ for the
maximum mass stable solution ($t_0=3.0$) and then increases.
The fractional mass defect, or the packing fraction, is 
$f = \Delta/y_b = -0.036$.
$|f|$ is only 3.6\% and may appear small. However, if we compare
this value to a typical nuclear packing fraction, which is
less than 0.1\% \citep{fermi}, we will find it to be significant.
$\Delta < 0$ is a necessary condition for stability, but
not sufficient. $\Delta < 0$ still holds for $3<t_0<5$, but
these solutions correspond to unstable equilibria. 
When $\Delta < 0$,
solutions are for stable equilibria for $d\Delta/dt_0 \le 0$, while they
are for unstable equilibria for $d\Delta/dt_0 > 0$. 

The question of whether dark matter fermions can form a black hole would be
answered in terms of the maximum gravitational
mass, $M_{max}$ and the corresponding
radius $R_{max}$, as functions of the particle mass $\mu_0$ and 
statistical weight $g$. The results are:

\begin{eqnarray}
M_{max} &=&1.73\times 10^{51}\frac{1}{\sqrt{g}}
\left(\frac{\mu_{0}}{\mathrm{eV}}\right)^{-2}\hspace{1cm}(\mathrm{g}) \\
&=&8.70\times 10^{17}\frac{1}{\sqrt{g}}
\left(\frac{\mu _{0}}{\mathrm{eV}}\right)^{-2}
\hspace{1cm}(\mathrm{M}_{\odot }),
\label{Mmax}
\end{eqnarray}
and

\begin{eqnarray}
R_{max} & = & 1.10\times10^{24}\frac{1}{\sqrt{g}} 
\left(\frac{\mu_0}{\mathrm{eV}}\right)^{-2} \hspace{1cm} (\mathrm{cm}) \\
& = & 3.54\times10^2\frac{1}{\sqrt{g}} 
\left(\frac{\mu_0}{\mathrm{eV}}\right)^{-2} 
\hspace{1cm} (\mathrm{kpc}).
\end{eqnarray}

It should be noted that \citet{landau} define the maximum rest mass, 
$M_{rest}=1.037M_{max}$ as the maximum mass, which is the total
rest mass brought from infinity.
In the case of a star made of free neutrons, $\mu_0 = 939$ MeV and $g=2$.
Then $M_{max} = 0.70 \mathrm{M}_\odot$, and $R_{max} = 8.8$ km. 
It should also be noted that $R_{max}$ is sensitive to the actual outer
boundary condition in the numerical calculation, which in principle
should be $p=0$. In reality, integration must be stopped at
a finite value of $p$ and $r_b$ is somewhat sensitive to this finite $p$.
Physically speaking, it is unrealistic to assume the fully degenerate
equation of state to a small $p$ and the outer boundary condition is
not well defined. The total gravitational mass $u_b$ is not very sensitive
to the choice of the actual outer boundary condition $p$ and is well
defined. Although it is not explicitly stated in the original paper,
\citet{OV39} were probably aware of this limitation
in the applicability of the equation of state, if we judge
from the carefully chosen title ``On Massive Neutron Cores''.
We also admit that without this boundary condition, we cannot
use the Schwarzschild's exterior solution outside the boundary
and the formulation becomes more complicated.

Before concluding this section, we comment on the nonrelativistic
limit. For the nonrelativistic limit, it is expected that between
the mass, $M$, and the size, $R$, of an FDFS,
the relation,

\begin{equation}
MR^3 = const.
\end{equation}

should hold \citep{landau}. 

For $t_0 < 0.5$, $u_b r_b^3$ is indeed nearly constant and 

\begin{equation}
u_b r_b^3 = 0.135,
\end{equation}

within 4\% precision.
In physical units,

\begin{eqnarray}
M R^3 & = & (u_b b)(r_b a)^3 \nonumber \\
     & = & 1.4\times10^{124}
\frac{1}{g^2}\left(\frac{\mu_0}{\rm eV}\right)^{-8} 
\hskip 10mm ({\rm g}\cdot{\rm cm}^3) \\
     & = & 2.3\times10^{26}
\frac{1}{g^2}\left(\frac{\mu_0}{\rm eV}\right)^{-8} \hskip 10mm 
({\rm M_\odot}\cdot{\rm kpc}^3). \label{mr3}
\end{eqnarray}

Nonrelativistic solutions are similar and in Fig.\ref{norm}, the profiles
of
the normalized mass density and 3D encircled mass are plotted.
In the next section, we will find the behavior of the 3D encircled mass profile
interesting and the logarithmic profile of the normalized 3D encircled
mass is given in Fig.\ref{lognorm}.

\section{Can degenerate fermions be
the dominant dark matter in the
cluster of galaxies, A1689?}

Recently \citet{TB05a,TB05b} 
reported a mass column density profile 
of the cluster of galaxies, 
A1689, obtained from gravitational lensing.
One of the important properties of the profile is that
it has a flat top.
We propose that this flat-top column density profile 
might be explained by the effects of degeneracy pressure of
fermionic dark matter. Here we analyze this proposal.

First we briefly introduce the main results of \citet{TB05a,TB05b}.
In their analysis, 1$^{\prime}$ corresponds to 129 kpc $h^{-1}$.
In \citet{TB05a}, the central 250 kpc $h^{-1}$ in radius
of multi-color HST/ACS images were analyzed.
The mass column density profile, $\Sigma(r)$,
is not expressed as a single power law
of radius.
The mass column density profile flattens toward the center with
a mean slope of dlog$\Sigma$/dlogr $\approx -0.55$ within 
$r<$250 kpc $h^{-1}$. Inside the Einstein radius 
($\theta_E\approx 50^{\prime\prime}$), they obtained the slope
of $\approx -0.3$ from the ratio between 
$\theta_E$ and the radius of the radial critical curve,
$\theta_r\approx 17^{\prime\prime}$. 
They fit their results with an inner region of an NFW profile
\citep{NFW}
with a relatively high concentration, $C_{vir} = 8.2$.

The mass column density, $\Sigma(r)$, is the integral
of the volume density, $\rho(r)$, 
along the line of sight over the entire cluster scale
of Mpc. In order to study the possibility of fermion degeneracy
near the center of the cluster,
we need information on the volume density, $\rho(r)$, instead of
the column density, $\Sigma(r)$.
\citet{TB05b} present the weak-lensing analysis of
the wide field data obtained by Subaru and obtained 
the column density profile at $r<2$ Mpc $h^{-1}$.
They fit the combined profile of HST/ACS and Subaru with
an NFW profile with a very high concentration,  $C_{vir} = 13.7$,
significantly larger than theoretically expected value of $C_{vir} \approx 4$.
They also fit the same observed column density profile with
a power law profile with a core. 
They give this result in terms of the angular radius dependence 
of the convergence, $\kappa$, as

\begin{equation}
\kappa \propto (\theta + \theta_C)^{-n}.
\label{kappa}
\end{equation}

$\theta_C = 1.65^{\prime}$ and $n = 3.16$ give the best fit although
$\theta_C$ and $n = 3.16$ are mutually dependent and a finite range
of the combination ($\theta_C$,$n$) gives equally good fits.
In terms of $\chi^2$ and the degrees of freedom, this
core power law profile fits the observation better than
the best-fit NFW profile and we use this profile for further discussion.
Although \citet{TB05b} do not claim so explicitly, the two facts that
the best-fit NFW profile shows a much higher concentration than
the value predicted by the CDM cosmology and the phenomenological
profile, Eq.(\ref{kappa}), fits better than the best-fit NFW profile,
indicate some serious contradiction to the CDM cosmology.

We start
our analysis from 
this core power-law profile, (\ref{kappa}), for further discussion.
We convert (\ref{kappa}) to a column density profile, $\Sigma(r)$,
in physical units of length and mass using the relations,
$\kappa = \Sigma/\Sigma_{crit}$, $\Sigma_{crit} \approx$ 0.95
${\rm g}\cdot{\rm cm}^{-2}$, and the normalization of 2D encircled
mass inside
the Einstein radius, $r_E$,
$\int_0^{r_E} \Sigma(r) 2\pi r dr = \Sigma_{crit}\cdot \pi r_E^2$.
The result is expressed as

\begin{equation}
\Sigma(r) = 25.2 \cdot \left(r/r_E + 2.2\right)^{-3.16}, 
\hskip 1cm ({\rm g}\cdot{\rm cm}^{-2})
\label{cpl}
\end{equation}

where $r_E = 97$ kpc $h^{-1}$
corresponds to $\theta_E=45^{\prime\prime}$,
the value used in \citet{TB05b}. 
The 2D encircled mass, $M_2(r) = \int \Sigma(r) 2 \pi r dr$, is
analytically obtained and  $M_2(r) = 1.3\times 10^{14} h^{-2} M_\odot$ and
$1.1\times 10^{15}h^{-2} M_\odot$ respectively for $r = r_E$ and $r=\infty$.
Therefore a high concentration of the mass is expected on
the scale of $r_E$. By assuming
spherical symmetry, we wish to obtain the volume density $\rho(r)$ by
solving 

\begin{equation}
\Sigma(x) = \int \rho(\sqrt{x^2+z^2}) dz,
\end{equation}

but we were not able to obtain an analytic solution.
Instead, we assumed another power law with a core for $\rho(r)$
and obtained the best fit parameters. The range of integration
in $z$ is from $-$2Mpc $h^{-1}$ to $+$2Mpc $h^{-1}$. The result is

\begin{equation}
\rho(r) = 1.60\times10^{-23} \left(r/r_E + 1.28\right)^{-3.71} h.
\hskip 1cm ({\rm g}\cdot{\rm cm}^{-3})
\label{3density}
\end{equation}

Near the center, $\rho(r)= 6.4\times10^{-24} h$ and $7.5\times10^{-25} h$
 (g$\cdot$ cm$^{-3}$)
respectively at $r$ = 0 and $r_E$. As for the case with the column density
profile, the core radius and power-law index are mutually dependent
and a finite range of their combination gives equally good fits.
The above volume mass densities  may appear small, but the
degeneracy depends on the number density and de Broglie wavelength,
both of which are heavily dependent on the rest mass of the particle.
Before proceeding to the analysis by an FDFS, we first confirm
that eV-mass fermions can be degenerate at these low mass densities.
Since 1 eV corresponds to $1.8\times10^{-33}$ g, the number density, 
$N/V \approx 10^{11}$ cm$^{-3}$, the mean inter-particle spacing,
$(N/V)^{-1/3} \approx 2\times 10^{-4}$ cm. On the other hand,
the de Broglie wavelength for a 1 eV particle with a velocity $v$ is,
$\hbar/\mu_0 v = \lambda_{Compton}\cdot(c/v) = 2\times10^{-5}(c/v)$ cm. 
Therefore for nonrelativistic particles with $v<0.1c$, the condition 
for degeneracy, $(N/V)^{-1/3}< \lambda_{({\rm de \hskip 5pt Broglie})}$, 
is satisfied. 

More quantitative discussion is possible for a 3D encircled mass profile
given in Fig.\ref{m3}. For the purpose of an explicit comparison,
here we fix $h = 0.7$. The 3D encircled mass profile also predicts
the rotation curve profile as Fig.\ref{Vkep}, which can be compared
with observations of kinematics of the galaxies in the cluster
in the future.
The column density profile, Eq.(\ref{cpl}), was derived 
as a phenomenological formula without assuming any
background physics, and so was the volume density profile,
Eq.({\ref{3density}). Here we wish to explain the presence of
the core in the power-law profile, which causes
the flat-top nature of Eq.(\ref{cpl}), in terms of an FDFS.
The comparison must be made either in the volume density profiles
or in the 3D encircled mass profiles. Here we choose the latter,
because the observed 3D encircled mass profile is much less
sensitive to the actual choice of the combination
of the core radius and the power-law index in the original Eq.(\ref{cpl}).
As we mentioned above, the finite range of the parameter combinations
gives equally good fits to the observations. The basic reason for this
is that our observables are integrated quantities rather than
the local densities.
First, we compare the slopes of the 3D encircled mass profiles
of the observation (Fig.\ref{m3}) and our model (Fig.\ref{lognorm}).
The observed slope is 2.93 at $r<$10 kpc and 2.54 at 10 $<r<$ 100 kpc.
On the other hand, the slope of a nonrelativistic FDFS is
2.90 between $r=0.03r_b$ and $0.3r_b$ and becomes shallower at
$0.3 r_b < r < r_b$. Note that the slope 
of the 3D encircled mass profile for a constant volume density
is 3.0 and both the observed and model inner slopes of 2.9 are consistent
with the flat-top nature of the volume density profile.
Now we proceed to the physical scaling of the model in terms
of the mass and length. For this purpose, it is most convenient
to use the relation Eq.(\ref{mr3}) for the mass and size of
a nonrelativistic FDFS.
The Eq.(\ref{mr3}) can be rewritten as

\begin{equation}
\log M_{14} + 3 \log R_{2} = f(g,\mu_0),
\end{equation}

where $M_{14} = M/(10^{14}M_\odot)$, $R_2 = R/(100 {\rm kpc})$ 
and $f(g,\mu_0) = 6.36  - 2 \log g - 8 \log(\mu_0/{\rm eV})$. 
In Fig.\ref{M3f}, the possible combinations of $M$ and $R$ are
plotted for different values of $f(g,\mu_0)$. Intersections of
the dotted lines and the solid line (3D encircled mass profile of A1689)
correspond to solutions for representative values
of $f$. Actually $f$ is continuous and
the solutions are continuous.
The probable range is  $f=-6 \sim 2$.
For this range of $f$, the rest mass of a fermion ranges from
30 to 2 eV. The larger the rest mass, the smaller the degenerate structure.
The possible range of $g$ is limited. For a particle with spin 1/2,
$g$ is either 1 (Majorana particle) or 2 (Dirac particle).

The special case is the massive neutrinos with similar masses for which
effective $g = 3$ (Majorana) or $g = 6$ (Dirac).
Recent underground experiments have shown that the mass differences
among three species of neutrinos are smaller than $0.05$ eV \citep{shirai}.
In order for massive neutrinos to have masses greater than 1 eV, they must 
have similar masses (degenerate in mass).
The mean number density of relic neutrinos is cosmologically
fixed and there is the well known relation 
for the contribution of the relic neutrinos
to $\Omega$,

\begin{equation}
\Omega_\nu h^2 = \Sigma_i \mu_i / (93.5{\rm eV}),
\end{equation}

where $i=1\sim3$ for Majorana neutrinos and $i=1\sim6$ for Dirac neutrinos.
The maximum allowed neutrino mass can be estimated by setting
$\Omega_\nu = 0.3$ and $h=0.7$, to be 4.7 or 2.3 eV respectively
for Majorana or Dirac neutrinos. The former gives $f=0$ or the
degenerate mass of $10^{14}M_\odot$ in 100 kpc,
and the latter gives $f=2$ or that of $5\times10^{14}M_\odot$ in 300 kpc.
These are rather comfortable numbers for this cluster profile.
If we question the CDM cosmology based on the possible inconsistency
of the high concentration of the observed mass profile
with the CDM predictions mentioned above, 
we might as well revive massive neutrinos (hot dark matter)
as the candidate for the dominant dark matter.

\section{Mass hierarchy of black holes from that of neutrino through FDFS?}

In the previous section, we have studied a possibility that fully degenerate
fermions forms a huge mass concentration at the center of a cluster of
galaxies and we mainly examined the case for massive neutrinos.
Is this possibility really true? In order to answer this question,
we focus on the universality and the scalability of FDFS.
We try to extend this idea of FDFS to other neutrinos and fermions
with different masses. As we have already seen in 
Eq.(\ref{Mfermi},\ref{Mmax}),
the rest mass
of the fermion mostly
determines the characteristic mass scale of the structure; more
massive fermions form lighter structures.

The most extreme condensed structure is a black hole. We have already
known that there exist many black holes of several species in the Universe
\citep{variousBH}.
They are the most familiar stellar mass  
black holes ($\approx M_{\odot}$), 
giant black holes at the
center of a galaxy ($\approx 10^{7}M_{\odot }$), 
and the intermediate mass black holes ($\approx 10^{3}M_{\odot }$).
Although these massive black holes are actively studied based
on the bottom-up scenarios that they are formed by the coalescence of the
stellar sized black holes, we try to propose yet another scenario
based on the context of FDFS. The most prominent property of those black
holes are that they appear to have a hierarchy in mass range. 
Most of the black
holes are classified in the above three types and those in other mass ranges
is rare. Therefore it would be natural to suspect any definite mechanism to
construct such hierarchy from the fundamental level. Our hypothesis is that
such fundamental mechanism is the mass hierarchy in fermions or possibly
neutrinos. If those black holes are formed by the overweight FDFSs, we can
estimate the masses of those neutrinos as in Table 2.

The corresponding masses of fermions, except $\nu _{e,\mu ,\tau }$,  through
Eq.(\ref{Mfermi}) are far heavier than eV and
we may identify those fermions
as more massive sterile neutrinos. 
Lighter neutrino of mass $10^{-1}-10^{-3}$ eV
would yield  FDFSs much extended dilute structures whose sizes well exceed
the Horizon size. In the above, we have approximate values.
However, the lower limit of the black hole mass within a class
yields the precise value of the corresponding neutrino mass. 
The existence of such a lower limit would be the key ingredient
of the FDFS model. 

\section{Conclusions and Discussions}

We have examined the possible structures formed by the fully degenerate 
self-gravitating fermions (FDFS) at various scales, such as
the mass concentration of
a cluster of galaxies, the giant black
holes at the center of a galaxy and the intermediate black holes in
galaxies. As an order
estimation, their characteristic masses directly reflect the constituent
fermion masses through the simple relation, 
$M_{fermi}=G^{-3/4}m^{-2}=m_{pl}^{3}/m^{2}$.
For the purpose of a quantitative analysis,
exact masses and detailed mass density profiles of FDFSs
were examined from nonrelativistic to relativistic situation,
using the formalism of Oppenheimer and Volkoff.

These results were applied to
the cluster of galaxies, A1689, whose mass distribution
has been observationally obtained.
We converted the observed column density profile to a volume density
profile assuming spherical symmetry, and compared
the observed and model 3D encircled mass profiles.
We found that the flat-top nature of the observed profile is
reproduced by the model and the particle mass range is between 2 eV and
30 eV depending on the actual scale of the degenerate structure.

For about black holes, our scenario will provide alternative mechanism of
the black hole formation. Most
of the present theories assume the coalescence of
stellar mass black holes in the gravitational potential. Such processes
seem to be quite complex compared to the FDFS
scenario, and therefore it would be
much difficult to explain, for example,
the observed universal relation between the black
hole mass at the center of a galaxy and the bulge mass: $M_{BH}/M_{bul}
\approx 0.002.$ This point will be discussed
in detail in our future work.







\acknowledgments
We thank Nobuo Arimoto for comments on the manuscript.
This study was motivated by a very interesting talk given by Tom
Broadhurst in 2004 at NAOJ, Mitaka, Japan.

\begin{figure}[tbp]
\includegraphics[angle=-90]{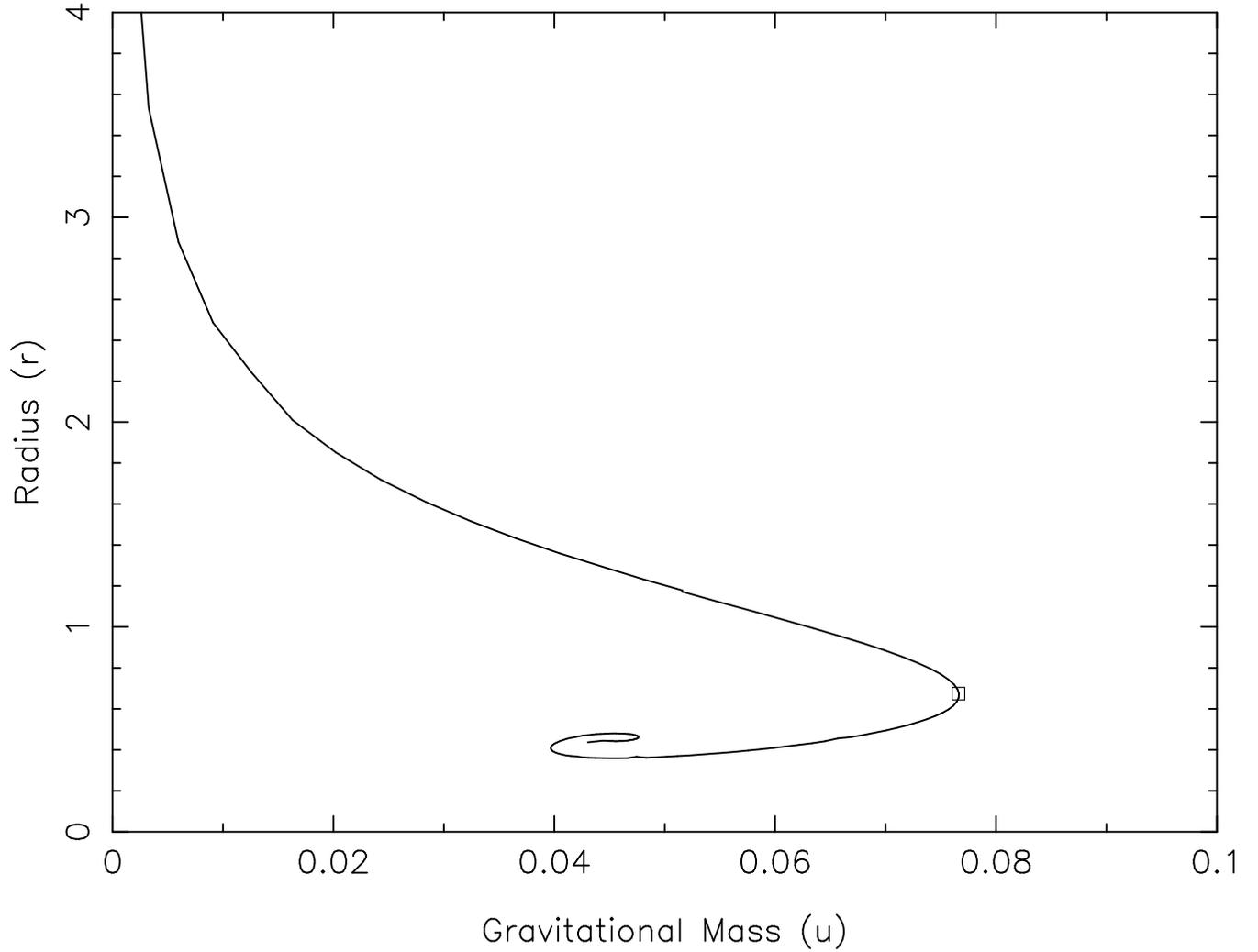}
\caption{Relation between the dimensionless gravitational mass $u$ and the
dimensionless radius $r$ of equilibrium configuration for various central
Fermi velocity ($v_F$). For $u < 0.04$, there is only one $r$, while $0.04 <
u < 0.0766$, there are multiple values of $r$. Among them, however, only the
configuration with the largest $r$ is the stable
equilibrium. The maximum dimensionless mass $u = 0.0766$ is the mass limit
before collapsing to a black hole (square). }
\label{fig1}
\end{figure}

\clearpage

\begin{figure}[tbp]
\includegraphics[angle=-90]{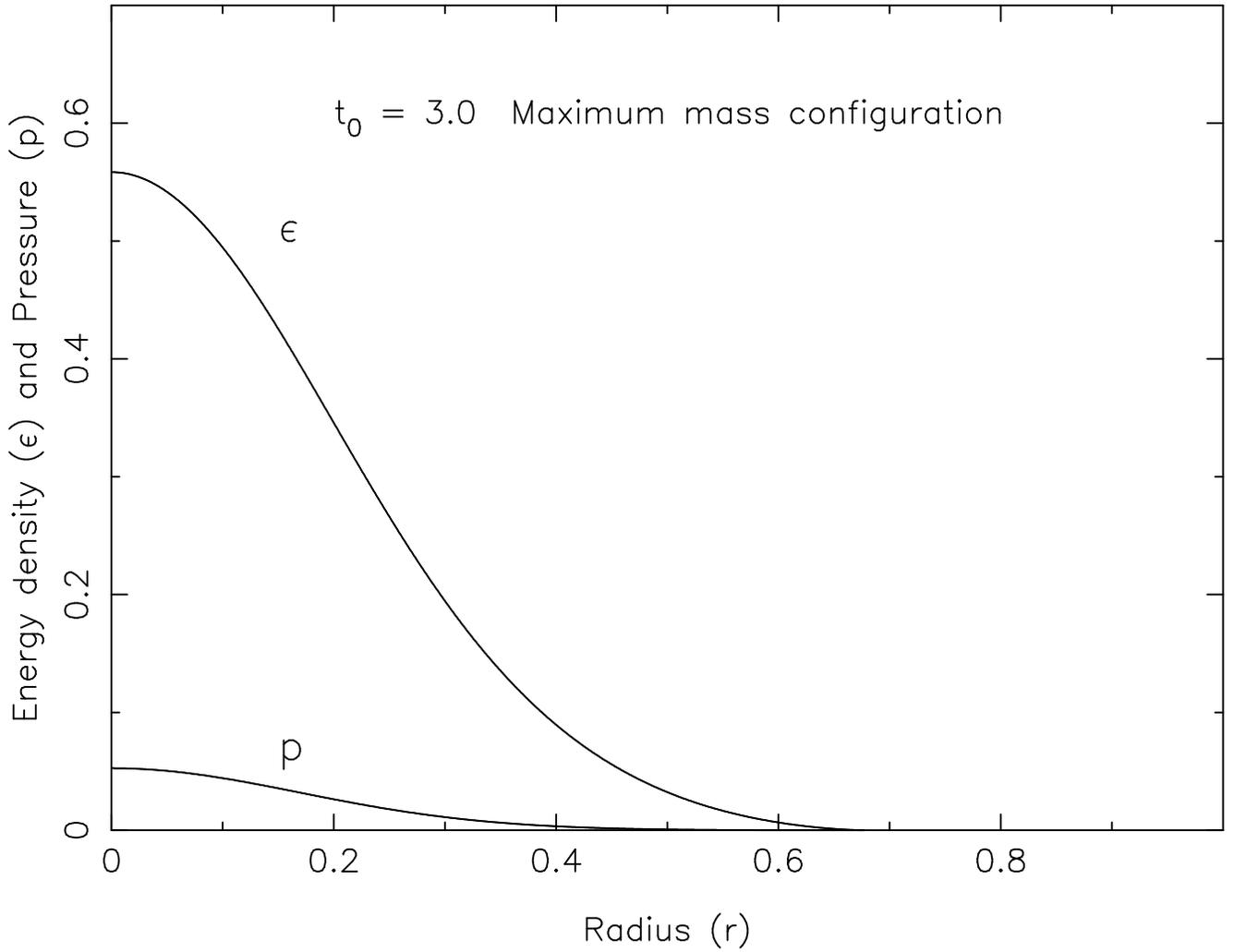}
\caption{Dimensionless energy density ($\epsilon$)
and pressure ($p$)
as functions of dimensionless radius for $t_0=3.0$.
This is the maximum mass stable configuration.
$p$ is small compared to $\epsilon$.
Degenerate particles are moderately relativistic and
the gravity is quasi-Newtonian.}
\label{fig2}
\end{figure}
\clearpage

\begin{figure}[tbp]
\includegraphics[angle=-90]{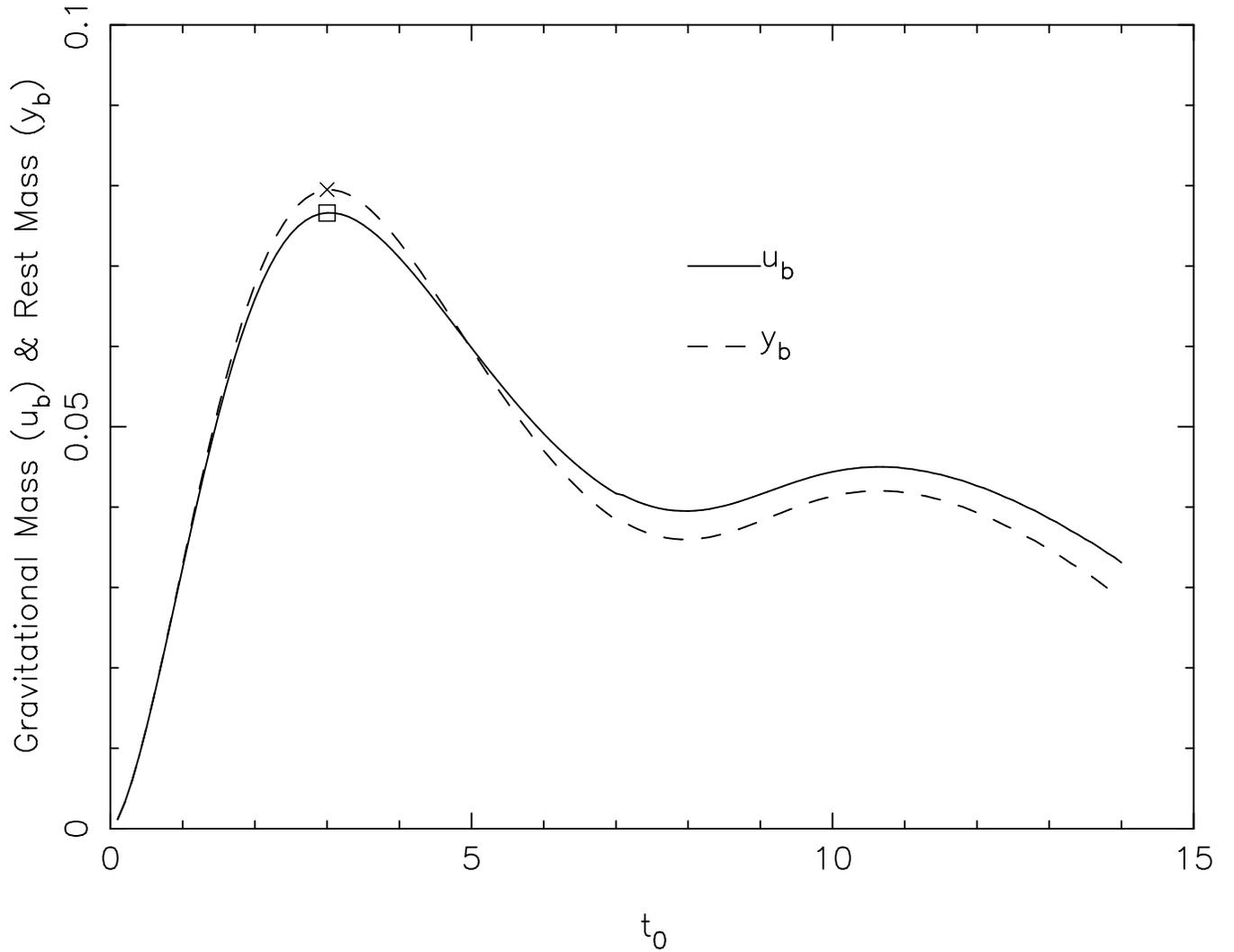}
\caption{Dimensionless gravitational mass ($u_b$)
and total rest mass ($y_b$) plotted as functions
of $t_0$.
The gravitational mass defect, $\Delta=u_b-y_b$, 
is negative for $t_0<3$ and takes
the minimum value for the maximum mass stable configuration ($t_0=3$).
}
\label{fig3}
\end{figure}

\clearpage


\begin{figure}[tbp]
\includegraphics[angle=-90]{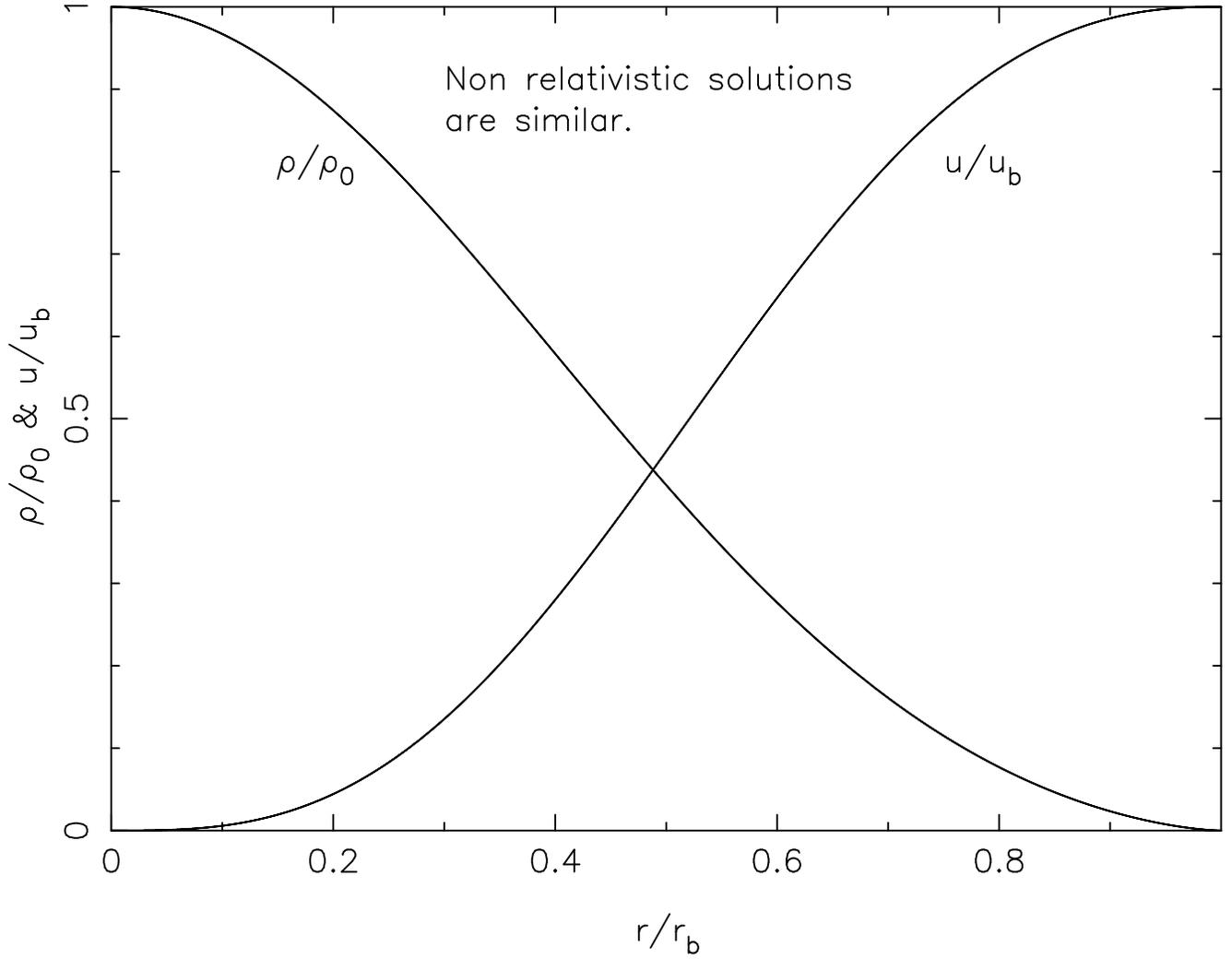}
\caption{Normalized density and encircled mass profiles for
a nonrelativistic FDFS. Nonrelativistic solutions are similar.}
\label{norm}
\end{figure}

\clearpage

\begin{figure}[tbp]
\includegraphics[angle=-90]{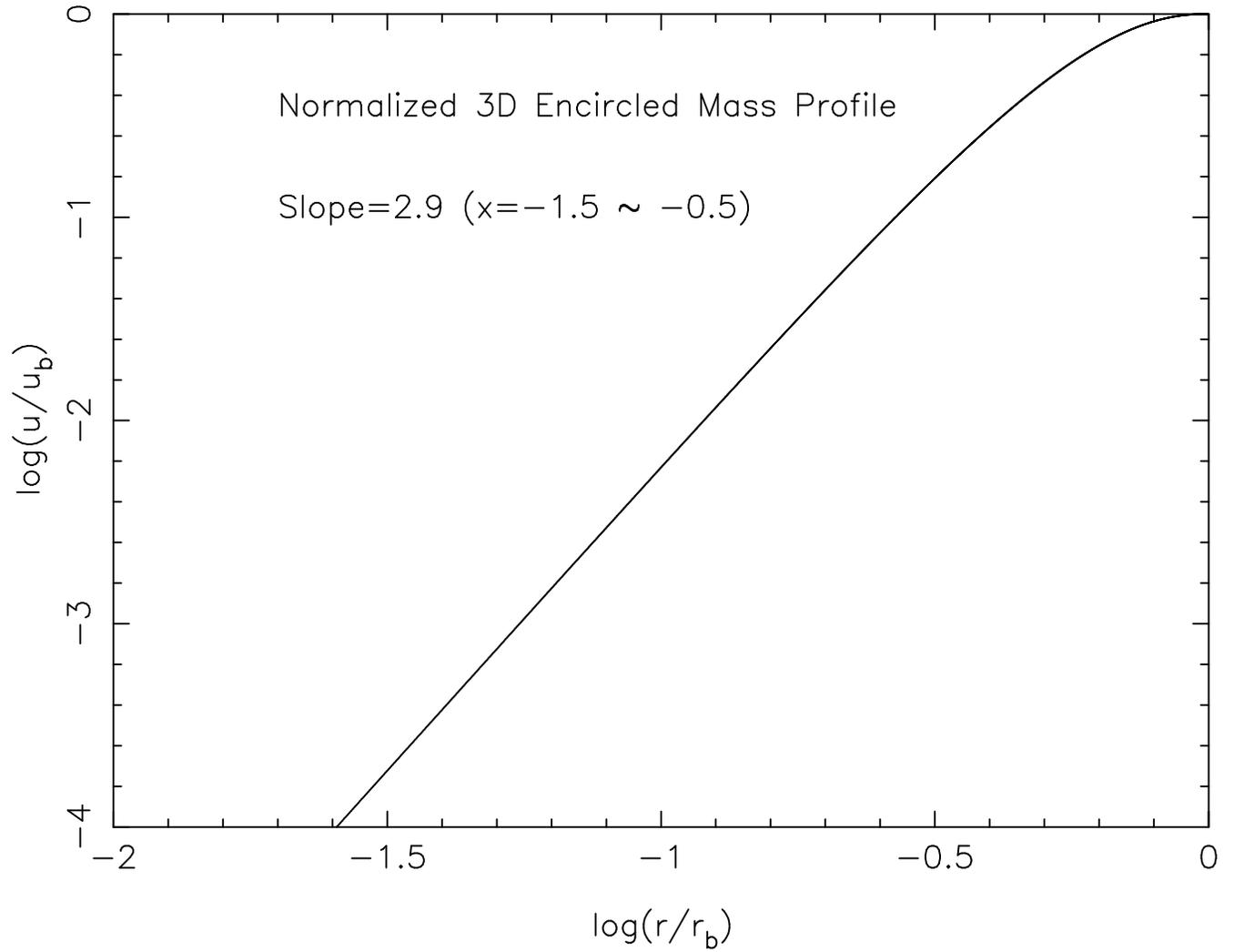}
\caption{Normalized 3D encircled mass profile of
a nonrelativistic FDFS in logarithmic scale.
The slope of the profile is 2.9 for $\log(r/r_b) = -1.5 \sim -0.5$.
}
\label{lognorm}
\end{figure}

\clearpage

\begin{figure}[tbp]
\includegraphics[angle=-90]{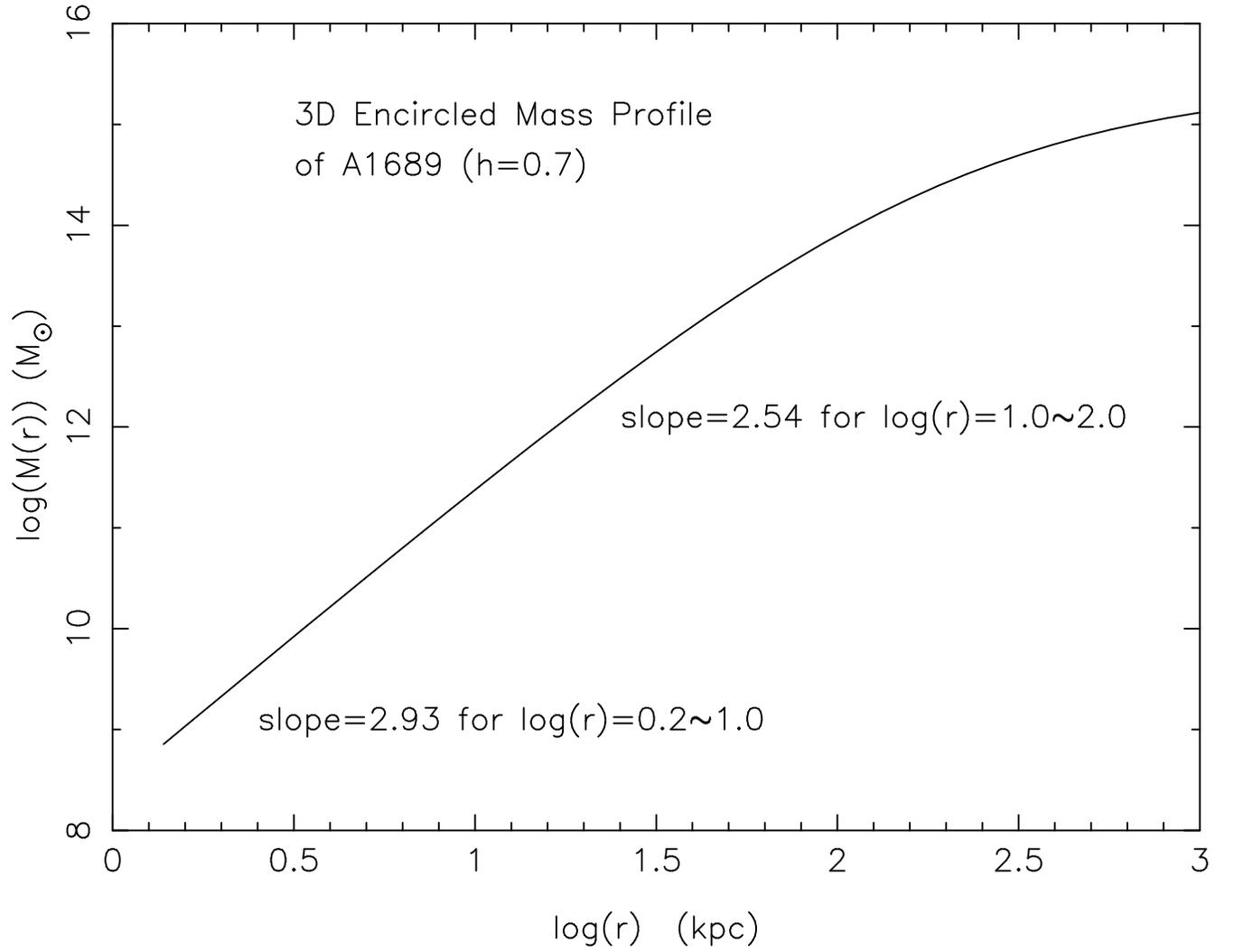}
\caption{3D encircled mass profile of A1689 obtained
from the best-fit volume density profile. h=0.7 is assumed.
}
\label{m3}
\end{figure}

\clearpage

\begin{figure}[tbp]
\includegraphics[angle=-90]{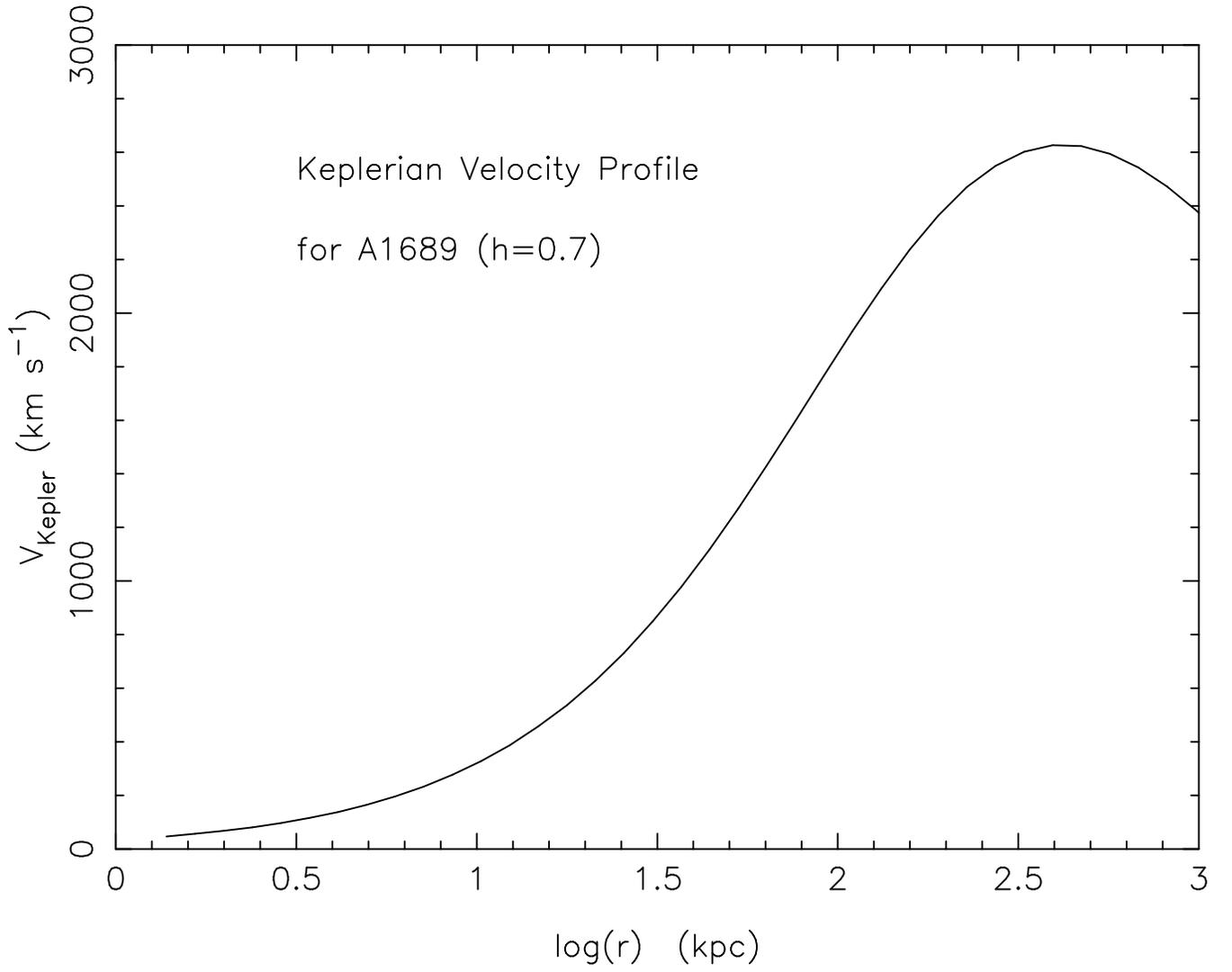}
\caption{Predicted rotation curve 
of A1689 estimated from the 3D encircled mass profile. h=0.7 is assumed.
}
\label{Vkep}
\end{figure}

\clearpage

\begin{figure}[tbp]
\includegraphics[angle=-90]{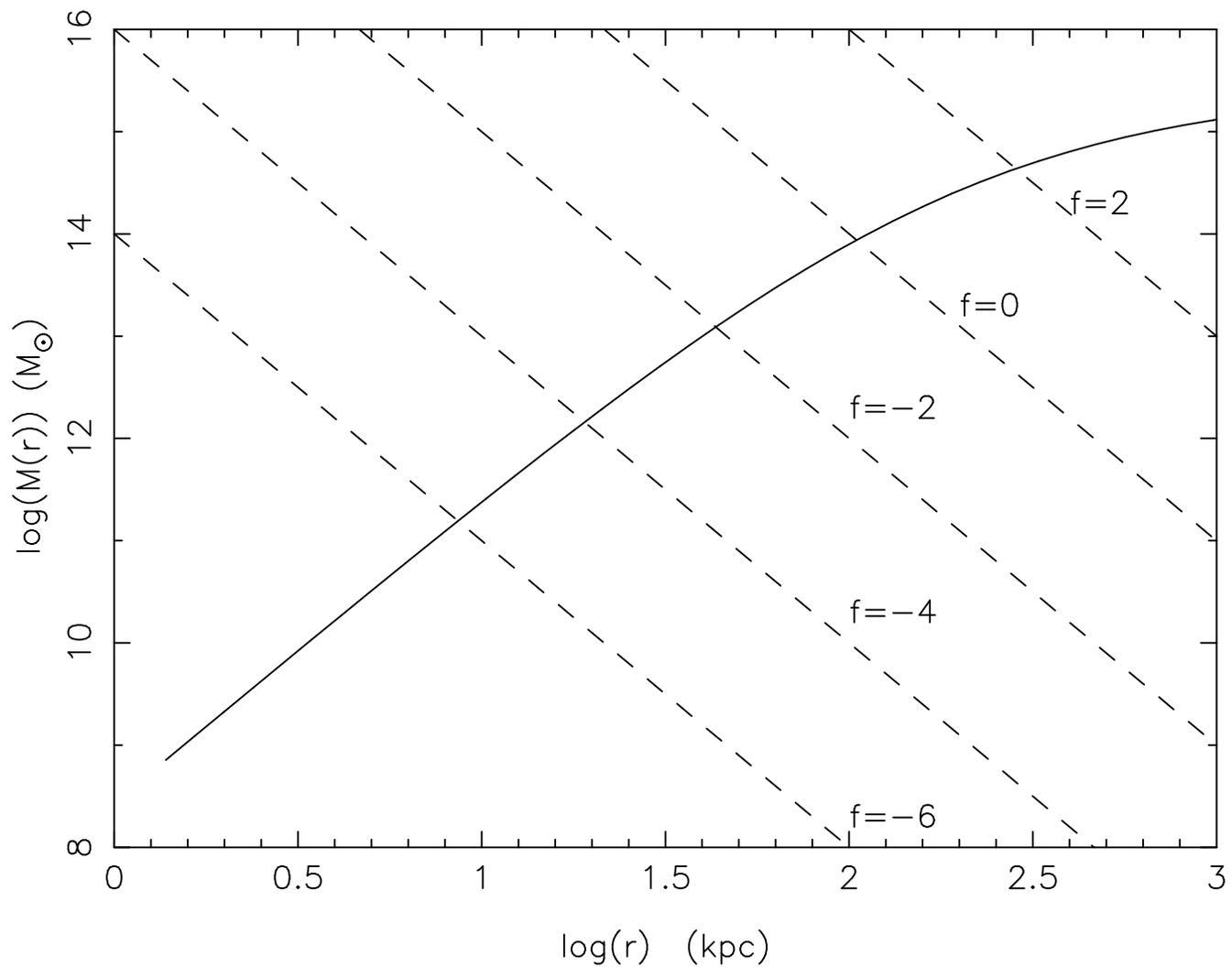}
\caption{Allowed combinations of the mass and size of an FDFS
for different values of 
$f(g,\mu_0) = 6.36 - 2 \log g - 8 \log(\mu_0/{\rm eV})$ (dotted lines). 
The solid line represents the 3D encircled mass profile of A1689.
Intersections between the solid and dotted lines correspond to
possible degenerate structures.
}
\label{M3f}
\end{figure}

\clearpage



\begin{table}
\begin{center}
\caption{Solutions for various $t_0$.\label{tbl-1}}
\vskip 3mm
\begin{tabular}{ccccc}
\tableline\tableline
$t_0$ & $v_{F0}/c$ & $u_b$ & $y_b$ & $r$ \\
\tableline
      & Fermi Vel. & Grav. mass & Rest Mass & Radius \\
\tableline
     0.1 &   0.025 &  0.0012 &  0.0012 &   4.934\\
     0.2 &   0.050 &  0.0033 &  0.0033 &   3.484\\
     0.3 &   0.075 &  0.0060 &  0.0060 &   2.838\\
     0.4 &   0.100 &  0.0091 &  0.0091 &   2.450\\
     0.5 &   0.124 &  0.0126 &  0.0126 &   2.183\\
     1.0 &   0.245 &  0.0325 &  0.0328 &   1.495\\
     1.5 &   0.358 &  0.0516 &  0.0525 &   1.161\\
     2.0 &   0.462 &  0.0659 &  0.0677 &   0.943\\
     2.5 &   0.555 &  0.0740 &  0.0766 &   0.784\\
     3.0\tablenotemark{a} &   0.635 &  0.0766 &  0.0795 &   0.663\\
     4.0 &   0.762 &  0.0710 &  0.0730 &   0.493\\
     5.0 &   0.848 &  0.0598 &  0.0597 &   0.391\\
     6.0 &   0.905 &  0.0491 &  0.0470 &   0.334\\
     7.0 &   0.941 &  0.0420 &  0.0387 &   0.343\\
     8.0 &   0.964 &  0.0395 &  0.0360 &   0.364\\
     9.0 &   0.978 &  0.0416 &  0.0383 &   0.382\\
    10.0 &   0.987 &  0.0444 &  0.0414 &   0.367\\
    11.0 &   0.992 &  0.0471 &  0.0443 &   0.399\\
    12.0 &   0.995 &  0.0463 &  0.0435 &   0.362\\
    13.0 &   0.997 &  0.0442 &  0.0411 &   0.324\\
    14.0 &   0.998 &  0.0411 &  0.0376 &   0.285\\
\tableline
\end{tabular}

\tablenotetext{a}{This solution corresponds to the maximum mass 
stable configuration. The gravitational mass defect, $u_b - y_b$,
takes the minimum.}

\end{center}
\end{table}

\clearpage

\begin{table}
\begin{center}
\caption{Neutrino masses and possible hierarchy \label{tbl-2}}
\vskip 3mm
\begin{tabular}{ccccc}
\tableline\tableline
& $m_{\nu }/{\rm eV}\leq $ & $M_{fermi}/M_{\odot }$ & $R_{fermi}$ & FDFS \\
\tableline 
$\nu _{e,\mu ,\tau }$ & $2.3$ & $9.8\times 10^{16}$ & 42 kpc & the center
of a cluster \\ 
$\nu _{1}$ & $0.18\times 10^{6}$ & $2.5\times 10^{7}$ & $460R_{\odot }$ & 
giant BH at center of galaxy \\ 
$\nu _{2}$ & $18.2\times 10^{6}$ & $2.7\times 10^{3}$ & $0.050R_{\odot }$ & 
intermediate BH in a galaxy \\
\tableline
\end{tabular}

\end{center}
\end{table}


\begin{thebibliography}

\bibitem[Bili\'c, Munyaneza \& Viollier(1999)]{bilic99} 
Bili\`c, N., Munyaneza, F. \&
Viollier, R. D. 1999, Phys. Rev. D, 59, 024003

\bibitem[Bili\'c, Tupper \& Viollier(2003)]{bilic03} 
Bili\`c, N., Tupper, G. B. \&
Viollier, R. D. 2003, astro-ph/0310294

\bibitem[Broadhurst et al.(2005a)]{TB05a}  Broadhurst et al., 2005, ApJ,
621, 53

\bibitem[Broadhurst et al.(2005b)]{TB05b}  Broadhurst et al., 2005, ApJ,
619, L143 

\bibitem[Fermi(1950)]{fermi} Fermi. E. 1950,
notes compiled by Orear, J, Rosenfeld \& Schluter,
R. A., Nuclear Physics (a course given by Enrico Fermi), Chicago

\bibitem[Gebhardt et al.(2002)]{variousBH}  Gebhardt, K., Rich, R. M. \& Ho,
L. C. 2002, ApJL, 578, 41


\bibitem[Kaup(1968)]{kaup} Kaup, D. J., 1968, Phys. Rev., 172, 1331

\bibitem[Landau \& Lifshitz(1980)]{landau}  Landau \& Lifshitz, 1980,
Statistical Physics, 3rd Ed.

\bibitem[Oppenheimer \& Volkoff(1939)]{OV39}  Oppenheimer, J. R. and
Volkoff, G. M. 1939, Phys. Rev. 55, 374


\bibitem[Navarro, Frenk \& White(1996)]{NFW} Navarro, J. F.,
Frenk, C. S., \& White, D. M. 1996, \apj, 462, 563

\bibitem[Shirai(2005)]{shirai}  Shirai, J. 2005, Nucl. Phys. B (Proc.
Suppl.), 144, 286

\bibitem[Tolman(1934)]{RT34}  Tolman, R. C., 1934, Relativity,
Thermodynamics and Cosmology, Oxford
\end{thebibliography}
\end{document}